%% file: main.tex
\documentclass[conf]{new-aiaa}
\usepackage[utf8]{inputenc}

\usepackage{graphicx}
\usepackage{amsmath}
\usepackage[version=4]{mhchem}
\usepackage{siunitx}
\usepackage{longtable,tabularx}
\usepackage{tikz}
\usepackage{booktabs}
\usepackage{graphicx}
\usepackage{pgfplots}
\pgfplotsset{compat=newest}
\usepgfplotslibrary{groupplots}

\usetikzlibrary{arrows.meta}
\usetikzlibrary{decorations.markings}

\usepackage{tabularx}
\usepackage{array}

\newcolumntype{Y}{>{\centering\arraybackslash}X}

\title{Pivot-Only Azimuthal Control and Attitude Estimation of Balloon-borne Payloads}

\author{Philippe Voyer\footnote{PhD Student, Department of Mechanical \& Aerospace Engineering, Princeton University.}, Simon Tartakovsky\footnote{PhD Candidate, Department of Physics, Princeton University.}, Steven J.~Benton\footnote{Senior Research Scholar, Department of Physics, Princeton University.}, and William C.~Jones\footnote{Professor, Department of Physics, Princeton University.}}
\affil{Princeton University, NJ, 08544, USA}

\begin{document}

\maketitle

\begin{abstract}
This paper presents an attitude estimation and yaw-rate control framework for balloon-borne payloads using pivot-only actuation, motivated by the Taurus experiment. Taurus is a long-duration balloon instrument designed for rapid azimuthal scanning at approximately $30^\circ$/s using a motorized pivot at the flight-train connection, without a reaction wheel. We model the gondola as a rigid body subject to realistic disturbances and sensing limitations, and implement a Multiplicative Extended Kalman Filter (MEKF) that estimates attitude and gyroscope bias by fusing inertial and vector-camera measurements. A simple PI controller uses the estimated states to regulate yaw rate. Numerical simulations incorporating representative disturbance and measurement noise levels are used to evaluate closed-loop control performance and MEKF behavior under flight-like conditions. Experimental tests on the Taurus gondola validate the pivot-only approach, demonstrating stable high-rate tracking under realistic hardware constraints. The close agreement between simulation and experiment indicates that the simplified rigid-body model captures the dominant dynamics relevant for controller design and integrated estimation-and-control development.
\end{abstract}

\section*{Nomenclature}

{\renewcommand\arraystretch{1.0}
\noindent\begin{longtable*}{@{}l @{\quad=\quad} l@{}}

$\mathbf{a}$ & unit rotation axis \\
$\mathbf{b}$ & gyro bias \\
$\delta\hat{\mathbf{b}}$ & bias error estimate \\
$c_z$ & Coulomb torque level \\
$\mathbf{c}$ & Coulomb friction vector \\
$\mathbf{C}$ & rotation matrix (e.g., $\mathbf{C}_{bi}$) \\
$\mathbf{D}$ & damping matrix \\
$e_z$ & error between desired and actual angular velocity \\
$f_{\mathrm{sc},j}$ & star camera sampling rate \\
$\mathbf{F}, \mathbf{G}$ & linearized error-state transition matrices \\ 
$\mathcal{F}$ & reference frame \\
$\mathbf{g}$ & gravity vector (in inertial or body frame) \\
$\mathbf{H}$ & linearized measurement matrix \\
$\mathbf{I}$ & payload inertia matrix \\
$k$ & stiffness coefficient \\
$k_p, k_i$ & proportional and integral control gains \\
$\mathbf{K}$ & Kalman gain \\
$L$ & pivot-to-center-of-mass (COM) distance \\
$m$ & mass of rigid body \\
$\mathcal{N}$ & normal distribution \\
$\mathbf{P}$ & error-state covariance matrix \\
$\mathbf{Q}$ & gyro noise covariance matrix, $\text{diag}(\boldsymbol\Sigma_g, \boldsymbol\Sigma_b)$ \\
$\mathbf{R}$ & camera measurement noise covariance matrix, $\text{diag}(\boldsymbol\Sigma_{n_1}, \boldsymbol\Sigma_{n_2})$ \\
$\mathbf{r}$ & vector measurement residual \\
$\mathbf{r}_{cm}$ & vector from pivot to the gondola COM \\
$\tau_{piv}$ & pivot motor torque \\
$\hat{\mathbf{x}}$ & global state estimate \\
$\delta\hat{\mathbf{x}}$ & error-state estimate, $[\delta\hat{\boldsymbol\theta} \ \delta\hat{\mathbf{b}}]^{\mathsf{T}}$ \\
$\mathbf{y}^b_j$ & vector measurement in body frame \\
$\boldsymbol\tau_i$ & torque vector $i$ \\
$\boldsymbol\tau_{ext}$ & external torque vector \\
$\boldsymbol\omega$ & body angular velocity vector \\
$\omega_\epsilon$ & Coulomb friction smoothing parameter \\
$\theta_{twist}$ & flight-train twist angle \\

$\delta\hat{\boldsymbol\theta}$ & attitude error estimate \\
$\delta\boldsymbol\xi \sim \mathcal{N}(0, \boldsymbol\Sigma_\xi)$ & multiplicative noise term \\
$\delta\mathbf{n}_j \sim \mathcal{N}(\mathbf{0}, \boldsymbol\Sigma_{n_j})$ & multiplicative measurement noise (camera) \\
$\boldsymbol\eta_g \sim \mathcal{N}(\mathbf{0}, \boldsymbol\Sigma_g)$ & gyro measurement noise \\
$\boldsymbol\eta_b \sim \mathcal{N}(\mathbf{0}, \boldsymbol\Sigma_b)$ & gyro bias noise (random walk) \\
$\boldsymbol\Psi_k$ & discrete-time exponential map \\
$\boldsymbol\Sigma_i$ & covariance matrices ($\xi$, $b$, $n_j$, $g$, $\tau$) \\
$\phi$ & rotation angle \\

\end{longtable*}}

\section{Introduction}
\begin{figure}[t]
  \centering
  \includegraphics[width=1\linewidth]{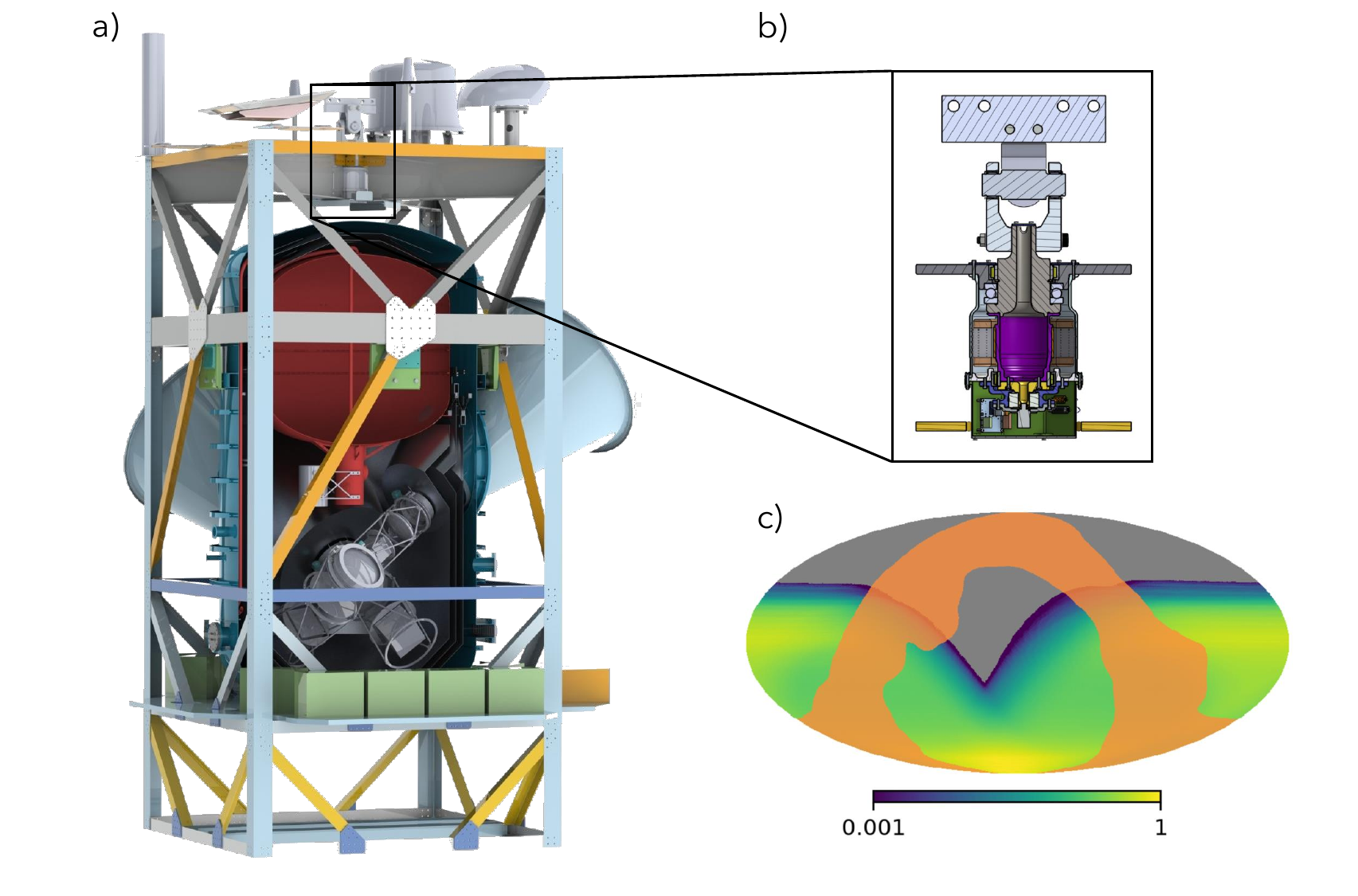}
  \caption{a) CAD model of the Taurus gondola with cross-sectional view of the instrument. b) A cross-sectional view of the pivot and universal joint assembly, with Parker K178200-8Y1-CE frameless servo motor in purple. c) Simulated normalized sky coverage of Taurus in equatorial coordinates for a late-March launch and one month of nightly observations, with the orange band indicating the Galactic plane and gray regions corresponding to unobserved areas \cite{Adler2024, May2024}.}
  \label{fig:1}
\end{figure}

\vspace{6pt}
Taurus is a balloon-borne cosmic microwave background (CMB) experiment designed to map E-mode polarization and refine measurements of early-universe reionization \cite{Adler2024, May2024, Tartakovsky2024}. Set to launch from New Zealand on a NASA Super-pressure Balloon (SPB), it is designed to spin in azimuth, or yaw, at $30^\circ\text{s}^{-1}$ while maintaining a fixed elevation of $35^\circ$. As the Earth rotates, the scan will naturally sweep across the sky, allowing Taurus to map approximately $70\%$ of the celestial sphere. The Taurus payload consists of three refracting telescopes mounted within a common liquid-helium cryostat and carried by a rigid gondola, as pictured in Fig.~\ref{fig:1}a.

This rapid azimuthal scan profile, combined with the dynamic stratospheric flight environment, introduces challenges for attitude determination and control. While the in-flight pointing control requirement for Taurus is modest, accurate pointing reconstruction is critical to meeting the science goals, particularly under highly dynamic conditions where conventional star-tracker solutions may be limited \cite{Du2024}. Taurus relies on a combination of a three-axis gyroscope, a three-axis magnetometer, and dual star cameras to track high-rate motion and estimate attitude and angular rate in real time. Yaw actuation is achieved through a motorized pivot at the connection to the flight train, with a universal joint that passively decouples the gondola from balloon-induced disturbances. The pivot and universal-joint assemblies, illustrated in Fig.~\ref{fig:1}b, build on designs with extensive flight heritage from SPIDER, SuperBIT, and EXCITE \cite{Gill2024, Shariff2014, Nagler2022}. The resulting azimuthal scan strategy was simulated in \cite{Adler2024, May2024}, with the corresponding hit-count map shown in Fig.~\ref{fig:1}c, indicating how the gondola’s attitude history projects onto the astronomical sky. Although often referred to as \textit{azimuthal} in the literature, the motion is more accurately \textit{yaw}, meaning rotation about the payload’s body axis rather than Earth-vertical; since they differ only by a small pendulation angle, the terms are used interchangeably here.

Conventional balloon-borne instruments typically use a combination of a pivot and a reaction wheel for yaw control and torque compensation \cite{Gill2024, Bernasconi2025, Pascale2008, Shariff2014, Cui2025}. In contrast, Taurus relies solely on a motorized pivot for actuation, eliminating the need for a reaction wheel to conserve mass and simplify the control architecture. Additionally, Taurus’ non-diagonal inertia forces rotation about a non-principal axis, creating inherent cross-axis coupling with pitch and roll. This introduces unique estimation and control challenges, as the system must maintain stability and achieve accurate attitude estimation despite limited actuation authority and coupling with the flight-train dynamics.

This paper presents a joint estimation and control framework for azimuthal (yaw) pointing of the Taurus payload using pivot-only actuation. A bias-aware multiplicative extended Kalman filter (MEKF) is implemented to estimate the system’s attitude and gyroscope bias by fusing inertial and optical measurements, and these estimated states are used in a feedback control loop to regulate the yaw rate during science operations. The payload is modeled as a rigid body subject to realistic disturbances, and simulation results are presented to evaluate the integrated performance of the estimator and controller under representative flight conditions. In addition to simulation results, preliminary experimental tests using a simple yaw-rate PI controller were conducted on the Taurus gondola to assess closed-loop yaw-rate tracking performance, overshoot, settling time, and steady-state jitter in a realistic hardware environment, demonstrating the feasibility of pivot-only actuation for high-rate scanning.

\section{Background}
\vspace{6pt}
This section outlines the modeling assumptions, notation, and theoretical background used to formulate the attitude estimation and control problem.

\subsection{Notation}
The attitude of one reference frame relative to another can be described by a rotation matrix, or direction cosine matrix (DCM), evolving on a nonlinear manifold, forming the special orthogonal group, denoted $\mathrm{SO}(3) = \left\{ \mathbf{C} \in \mathbb{R}^{3 \times 3} \,\middle|\, \mathbf{C}\mathbf{C}^\mathsf{T} = \mathbf{1},\ \det(\mathbf{C}) = +1 \right\}$, where $\mathbf{1}$ is the $3 \times 3$ identity matrix. The attitude of a body-fixed frame $\mathcal{F}_b$ with respect to an inertial frame $\mathcal{F}_i$ is fully described by the rotation matrix $\mathbf{C}_{bi} \in \mathrm{SO}(3)$, providing a complete (unique and global) parametrization of the rigid body's attitude. These rotation matrices can be further parametrized using standard conventions, namely Euler angles, unit quaternions or an axis-angle pair, each with distinct properties. In particular, Euler’s rotation theorem states that any rotation can be represented by a single angle $\phi$ about a unit axis $\mathbf{a}$. As such, the rotation matrix $\mathbf{C}_{bi}$ can be parametrized as
\begin{equation}
\mathbf{C}_{bi} = \mathbf{C}_{bi}(\mathbf{a}, \phi) = \exp{(-\mathbf{a}^\times \phi)}
\label{eq:axisangle}
\end{equation}
\noindent with negative sign accounting for the transformation from \( \mathcal{F}_i \) to \( \mathcal{F}_b \). The cross product operator $(\cdot)^\times$ is denoted as 
\begin{equation}
    \mathbf{v}^\times = \begin{bmatrix}
        0 & -v_3 & v_2 \\ v_3 & 0 & -v_1 \\ -v_2 & v_1 & 0
    \end{bmatrix}, \quad \forall\mathbf{v}=\begin{bmatrix}
        v_1\\v_2\\v_3
    \end{bmatrix} \in \mathbb{R}^3
\end{equation}
\noindent with inverse mapping 
\begin{equation}
    \mathbf{S}^\otimes = \begin{bmatrix}
        s_{11} & s_{12} & s_{13} \\ s_{21} & s_{22} & s_{23} \\ s_{31} & s_{32} & s_{33}
    \end{bmatrix}^\otimes=\frac{1}{2}\begin{bmatrix}
        s_{32}-s_{23}\\s_{13}-s_{31}\\s_{21}-s_{12}
    \end{bmatrix},\ \forall\mathbf{S} \in \mathbb{R}^{3\times 3}.
\end{equation}

\noindent The Taylor series expansion of the matrix exponential gives
\begin{equation}
\mathbf{C}_{bi} = \mathbf{1} - \mathbf{a}^\times \phi + \frac{(\mathbf{a}^\times)^2 \phi^2}{2!}
- \frac{(\mathbf{a}^\times)^3 \phi^3}{3!}
+ \frac{(\mathbf{a}^\times)^4 \phi^4}{4!} - \cdots
\label{eq:taylor}
\end{equation}
\noindent which recursively yields
\begin{equation}
\mathbf{C}_{bi} = \mathbf{1} + \sin\phi \cdot \mathbf{a}^\times + (1 - \cos\phi)  (\mathbf{a}^\times)^2,
\label{eq:Rodrigues}
\end{equation}
\noindent known as Rodrigues' rotation formula.

Mappings between these attitude parametrizations are available in literature and are left out for brevity. They will be used interchangeably as needed in this paper. We follow the general notation of \cite{Barfoot2024} and \cite{deRuiter2012}.

\subsection{System Dynamics}
It is common to model gimballed balloon-borne telescopes as rigid multibody chains \cite{Romualdez2018, Quadrelli2004, Kassarian2024}. In contrast, the Taurus gondola has a single motorized pivot connection and can be simplified as a single rigid body (Fig.~\ref{fig:2}). The attitude dynamics of a rigid body are described in continuous time by Euler's equations 
\begin{equation}
    \mathbf{I}\, \dot{\boldsymbol{\omega}}+\boldsymbol{\omega}^\times (\mathbf{I}\,\boldsymbol{\omega})=\boldsymbol{\tau}_{ext},
    \label{eq:Euler}
\end{equation}
\noindent where $\mathbf{I}=\mathbf{I}^\mathsf{T}>\mathbf{0}$ is the inertia matrix of the rigid body evaluated in the local frame, $\boldsymbol{\tau}_{ext}\in \mathbb{R}^3$ is the external torque applied on the rigid body, and $\boldsymbol{\omega} \in \mathbb{R}^3$ is the angular velocity of $\mathcal{F}_b$ relative to $\mathcal{F}_i$, expressed in $\mathcal{F}_b$, i.e., the angular velocity of the rigid body resolved in its local frame. We consider the various external torques acting on the gondola, including a gravity torque, a control torque from the pivot motor, a flight train torque, a damping term, and a Coulomb friction term, that is
\begin{equation}
    \boldsymbol \tau _{ext} = \boldsymbol \tau_g
+ \boldsymbol \tau_c + \boldsymbol \tau_{ft} +\boldsymbol\tau_d + \mathbf{c}+\boldsymbol \tau_{noise} \end{equation}
\noindent where
\begin{equation}
    \boldsymbol \tau_c=[0\ \ 0\ \ \tau_{piv}]^\mathsf{T}
\end{equation}
\noindent is the pivot motor control torque applied in yaw, and
\begin{equation}
    \boldsymbol\tau_g = \mathbf r_{cm}^\times (m\, \mathbf g_b) = \mathbf r_{cm}^\times (m\, \mathbf C_{bi}\,\mathbf g) 
\end{equation}

\noindent is the gravity torque at the pivot. We consider $\mathbf r_{cm} \in \mathbb{R}^{3} $ as the distance vector of length $L$ from the pivot to the center of mass (COM) of the gondola, expressed in the body frame, and $\mathbf g=[0\ \ 0\ \ -9.81]^\mathsf{T}\ \mathrm{m/s^2}$ is the inertial frame gravity vector, which is rotated into the body frame by $\mathbf{C}_{bi} \in \mathrm{SO}(3)$. The azimuthal flight train torque is commonly modeled as a torsional spring acting on the twist angle of the suspension, $-k\,\theta_{twist}$~\cite{Romualdez2018, Kassarian2024}, but is neglected in this first-pass analysis. A linear viscous damping term $\boldsymbol{\tau}_{d} = -\mathbf{D}\,\boldsymbol{\omega}$ is included to model energy dissipation due to internal friction and aerodynamic drag, and a constant Coulomb friction torque $\mathbf{c}$ is considered about the yaw axis. The Coulomb friction term, which was added to capture the nearly linear response observed in the experimental data shown in Fig.~\ref{fig:3}, is approximated as

\begin{equation}
    \mathbf{c}= \begin{bmatrix}
        0\\0\\-c_z\  \text{sign}(\omega_z)
    \end{bmatrix}\approx \begin{bmatrix}
        0\\0\\-c_z\  \text{tanh}(\omega_z/\omega_\epsilon)
    \end{bmatrix},
\end{equation}
\noindent where $c_z$ is the Coulomb torque level and $\omega_\epsilon$ is a smoothing tuning parameter. Finally, the term $\boldsymbol{\tau}_{noise} \sim \mathcal{N}(\mathbf{0}, \mathbf{\Sigma}_{\tau})$ represents a zero-mean, normally distributed process noise that accounts for unmodeled disturbances.

Next, rigid-body rotational kinematics are described by Poisson's kinematic equation
\begin{equation}
    \dot{\mathbf{C}}_{bi} = -\boldsymbol{\omega}^{\times} \mathbf{C}_{bi} .
    \label{eq:Poisson}
\end{equation}
\noindent Direct numerical integration of Eq. (\ref{eq:Poisson}) is possible, but does not preserve the orthogonality condition $\mathbf{C}_{bi} \mathbf{C}_{bi}^\mathsf{T} = \mathbf{1}$, leading to $\mathbf{C}_{bi} \notin \mathrm{SO}(3)$. Instead, in discrete time, the attitude is incrementally rotated at each time step such that 
\begin{equation}
    \mathbf{C}_{bi,k+1} = \boldsymbol{\Psi}_k\, \mathbf{C}_{bi,k},
    \label{eq:expmap}
\end{equation}
\noindent where $\boldsymbol{\Psi}_k \in \mathrm{SO}(3)$ is obtained from the axis–angle parametrization (Eq.~(\ref{eq:axisangle})), and is given by
\begin{equation}
    \boldsymbol{\Psi}_k = \exp(-\Delta t \cdot \boldsymbol{\omega}_{k}^\times).
\end{equation}

\noindent The exponential map $\boldsymbol{\Psi}_k$ preserves orthogonality, ensuring $\mathbf{C}_{bi,k} \in \mathrm{SO}(3)\ \forall k$. In practice, the exact matrix exponential is computed using Rodrigues' formula, Eq. (\ref{eq:Rodrigues}), to reduce computational cost. We can equivalently introduce noise in the signal using a noise rotation matrix multiplying Eq. (\ref{eq:expmap}), which for small angles can be approximated as the linearized Eq. (\ref{eq:taylor}), that is $ (\mathbf{1} - \delta\boldsymbol{\xi}_k^\times)$, where $\delta\boldsymbol{\xi}_k \sim \mathcal{N}(\mathbf{0}, \boldsymbol{\Sigma}_{\xi})$. Unlike additive noise, a multiplicative noise ensures $\mathbf{C}_{bi,k} \in \mathrm{SO}(3)$. As such, we have
\begin{equation}
    \mathbf{C}_{bi,k+1} \approx (\mathbf{1} - \delta\boldsymbol{\xi}_k^\times)\,\boldsymbol{\Psi}_k\, \mathbf{C}_{bi,k}.
    \label{eq:fullexpmap}
\end{equation}

\input{figures/fig2}

\subsection{Sensor Measurements}
It is assumed for numerical simulation that a body-mounted three-axis gyroscope provides measurements of the gondola's angular velocity, and a pair of dual star cameras provide attitude observations for absolute reference. Additional sensors, such as a three-axis magnetometer and a pivot encoder, are included on the payload hardware but are omitted from the simulation model.

\subsubsection{Gyroscope}
Assuming negligible structural flexure, the rate gyroscope measures
\begin{equation}
    \boldsymbol{\omega}_{b,k}^g = \boldsymbol{\omega}_{b,k} + \mathbf{b}_k + \boldsymbol{\eta}_{g,k},
    \label{eq:gyromodel}
\end{equation}
\noindent where $\boldsymbol{\omega}_{b,k}^g$ is the measured angular velocity, $\boldsymbol{\eta}_{g,k} \sim \mathcal{N}(\mathbf{0}, \boldsymbol{\Sigma}_g)$ is the gyro measurement noise, and $\mathbf{b}_k \in \mathbb{R}^3$ is a bias that evolves as a discrete-time random walk:
\begin{equation}
    \mathbf{b}_{k+1} = \mathbf{b}_k + \boldsymbol{\eta}_{b,k},
    \label{eq:rwbias}
\end{equation}
where $\boldsymbol{\eta}_{b,k} \sim \mathcal{N}(\mathbf{0}, \boldsymbol{\Sigma}_b)$. Sensor axial and rotational misalignments are neglected here, and $\boldsymbol\Sigma_g$, $\boldsymbol\Sigma_b$ are chosen to be representative of commercial off-the-shelf (COTS) gyroscopes.

 \subsubsection{Star Camera}
A pair of star cameras provides vector measurements of Taurus’s attitude. Under highly dynamic conditions and fixed exposure times, each camera may capture a streaked star field; however, for this analysis, each camera is assumed to output a single vector measurement. The measurement model treats each observation as a known unit vector in the inertial frame (i.e., right ascension and declination of a target) that is transformed into the body frame as
\begin{equation}
    \mathbf{y}_b^{\,j} = (\mathbf{1} - \delta\mathbf{n}_j^{\times})\,\mathbf{C}_{bi}\,\mathbf{y}_i^{\,j}, 
    \quad j = 1,2,
    \label{eq:vectormeas}
\end{equation}
\noindent
where $\mathbf{y}_i^{\,j} \in \mathbb{S}^2$ are known inertial reference vectors, with 
$\mathbb{S}^2 = \{\, \mathbf{r} \in \mathbb{R}^3 \mid \lVert \mathbf{r} \rVert_2 = 1 \,\}$.
The term $(\mathbf{1} - \delta\mathbf{n}_j^{\times})$ represents a small linearized rotation introducing measurement noise, 
where $\delta\mathbf{n}_j \sim \mathcal{N}(\mathbf{0}, \boldsymbol{\Sigma}_n)$. To uniquely determine the attitude, the two measurement vectors ($j=1,2$) must be linearly independent, i.e., not collinear. Each star camera is modeled as operating at a fixed sampling rate $f_{\mathrm{sc},j}$, providing discrete attitude measurements every $1/f_{\mathrm{sc},j}$ seconds.

\subsection{Attitude Estimation}
Accurate attitude knowledge is essential to meet the science objectives of the mission by ensuring precise pointing stability and enabling high-fidelity post-flight image reconstruction. Although real-time estimation is not required for control in Taurus, implementing an onboard filter facilitates performance assessment, supports in-flight monitoring, and provides a foundation for future autonomous operation. Attitude estimation on $\mathrm{SO}(3)$ requires preserving the structure of the rotation manifold, which is commonly done using the Multiplicative Extended Kalman Filter (MEKF) \cite{Crassidis2007}.

\subsubsection{Multiplicative Extended Kalman Filter (MEKF)}

The MEKF extends the classical EKF by preserving the $\mathrm{SO}(3)$ constraints on $\mathbf{C}_{bi}$. The general idea is to linearize only a small attitude error in the tangent space and apply the correction multiplicatively onto the global attitude estimate. Many MEKF frameworks in literature consider a unit quaternion parametrization of attitude, but the DCM is used directly here. We consider the global state $\hat{\mathbf{x}}_k = \{\,\hat{\mathbf{C}}_{bi,k}\ \ \hat{\mathbf{b}}_k\,\}$, where the hat symbol $(\hat{\cdot})$ denotes an estimated quantity. Note that $\hat{\mathbf{x}}_k$ is not expressed as a single column vector but as a composite state, with elements maintained, propagated, and corrected individually.
The corresponding error state is defined as $\delta\mathbf{x}_k = [\delta\boldsymbol{\theta}_k^\mathsf{T}\ \ \delta\mathbf{b}_k^\mathsf{T}]$, with the associated covariance $\mathbf{P}_k = \mathbb{E}[\delta\mathbf{x}_k\ \delta\mathbf{x}_k^\mathsf{T}] \in \mathbb{R}^{6\times6}$. Here, $\delta\boldsymbol{\theta}_k$ denotes the small rotation vector in the axis–angle representation describing the attitude error between the estimated and true orientations. One iteration of the filter proceeds as follows:\\

\noindent \textit{a) Propagation.}
At each time step, the gyroscope measurement $\boldsymbol{\omega}_{b,k}^g$ is bias-corrected to obtain the estimated body-rate,
\[
\tilde{\boldsymbol{\omega}}_{b,k} = \boldsymbol{\omega}_{b,k}^g - \hat{\mathbf b}_k.
\]
The attitude estimate is then propagated forward using the deterministic kinematics in Eq.~(\ref{eq:expmap}),
\begin{equation}
    \hat{\mathbf{C}}_{bi,k+1}^{-}
    = \exp(-\Delta t\,\tilde{\boldsymbol{\omega}}_{b,k}^{\times})
      \hat{\mathbf{C}}_{bi,k},
    \label{eq:mekf_C_propagate}
\end{equation}
where $(\cdot)^-$ and $(\cdot)^+$ denote the estimate prior and posterior, respectively.  
The gyro bias is propagated as constant with $\hat{\mathbf b}_{k+1}^{-} = \hat{\mathbf b}_{k}$, and the covariance of the $6\times6$ error state, $\mathbf{P}_k$, evolves according to
\begin{equation}
    \mathbf{P}_{k+1}^{-}
      = \mathbf{F}_k\,\mathbf{P}_k^+\,\mathbf{F}_k^{\mathsf{T}}
        + \mathbf{G}_k\,\mathbf{Q}\,\mathbf{G}_k^{\mathsf{T}},
\end{equation}
\noindent where $\mathbf{Q}\approx \operatorname{diag}(\boldsymbol{\Sigma}_g,\boldsymbol{\Sigma}_b)$ is the noise covariance matrix, and $\mathbf{F}_k$ and $\mathbf{G}_k$ are the linearized error state-transition matrices that evaluate to
\begin{equation}
    \mathbf{F}_k =
\begin{bmatrix}
\mathbf{1}-\Delta t\,\tilde{\boldsymbol\omega}_{b,k}^{\times} & -\Delta t\,\mathbf{1}\\[2pt]
\mathbf{0}_{3\times3} & \mathbf{1}
\end{bmatrix},
\qquad
\mathbf{G}_k =
\begin{bmatrix}
\mathbf{1} & \mathbf{0}_{3}\\
\mathbf{0}_{3} & \mathbf{1}
\end{bmatrix}.
\end{equation}
Both the global state $\hat{\mathbf{x}}_k$ and the error-state covariance $\mathbf{P}_k$ are updated at each time step $k$ corresponding to an available gyroscope measurement, which is commonly referred to as dead reckoning.\\

\noindent \textit{b) Measurement update.}
When inertial sensor measurements are available, they are incorporated through the measurement update step of the filter. In general, all available sensor models can be assembled within this framework, but in this analysis, only the star camera measurements are considered. For each star camera $j\in\{1,2\}$ with known inertial reference $\mathbf{y}_i^{\,j}$, the predicted body-frame vector is
\[
\hat{\mathbf{y}}_{b,k+1}^{\,j} = \hat{\mathbf{C}}_{bi,k+1}^{-}\,\mathbf{y}_i^{\,j}.
\]
The residual is formed by stacking the measurement differences,
\begin{equation}
    \mathbf{r}_{k+1} =
    \begin{bmatrix}
      \mathbf{y}_{b,k+1}^{\,1}-\hat{\mathbf{y}}_{b,k+1}^{\,1}\\
      \mathbf{y}_{b,k+1}^{\,2}-\hat{\mathbf{y}}_{b,k+1}^{\,2}
    \end{bmatrix},
\end{equation}
and linearizing about the current attitude estimate gives the measurement matrix
\begin{equation}
\mathbf{H}_{k+1} =
\begin{bmatrix}
 (\hat{\mathbf{y}}_{b,k+1}^{\,1})^{\times} & \mathbf{0}\\
 (\hat{\mathbf{y}}_{b,k+1}^{\,2})^{\times} & \mathbf{0}
\end{bmatrix}.
\end{equation}
The Kalman gain, correction vector, and covariance update are then
\begin{align}
\mathbf{K}_{k+1} &=
   \mathbf{P}_{k+1}^{-}\mathbf{H}_{k+1}^{\mathsf{T}}
   \bigl(\mathbf{H}_{k+1}\mathbf{P}_{k+1}^{-}\mathbf{H}_{k+1}^{\mathsf{T}}
         +\mathbf{R}\bigr)^{-1},\\[4pt]
\delta\mathbf{x}_{k+1} &= \mathbf{K}_{k+1}\mathbf{r}_{k+1},\\[4pt]
\mathbf{P}_{k+1}^{+} &=
   \bigl(\mathbf{1}-\mathbf{K}_{k+1}\mathbf{H}_{k+1}\bigr)
   \mathbf{P}_{k+1}^{-},
   \label{eq:mekf_P_update}
\end{align}
where $\mathbf{R}\approx \operatorname{diag}(\boldsymbol{\Sigma}_{n,1},\boldsymbol{\Sigma}_{n,2})$ collects the camera measurement noise covariances.\\

\noindent \textit{c) State update.}
Finally, the multiplicative correction is applied to the attitude estimate.  
By partitioning $\delta\mathbf{x}_{k+1}=[\delta\boldsymbol{\theta}_{k+1}^{\mathsf{T}}\;\;\delta\mathbf{b}_{k+1}^{\mathsf{T}}]^{\mathsf{T}}$,
\begin{align}
\hat{\mathbf{C}}_{bi,k+1}^{+} &=
   \exp(-\delta\boldsymbol{\theta}_{k+1}^{\times})
   \hat{\mathbf{C}}_{bi,k+1}^{-},\\[2pt]
\hat{\mathbf{b}}_{k+1}^{+} &=
   \hat{\mathbf b}_{k+1}^{-} + \delta\mathbf b_{k+1},
   \label{eq:est_updates}
\end{align}
where $\delta\boldsymbol{\theta}_{k+1}$ represents the small-angle attitude correction applied on the Lie group $\mathrm{SO}(3)$, and $\delta\mathbf b_{k+1}$ updates the gyro bias additively.

\subsection{Control}
\label{sec:control}
For the scope of this paper, a simple Proportional-Integral (PI) controller is implemented to actuate the pivot motor. We define 
\begin{equation}
    \tau_{piv} = k_p \ e_z + k_i  \int e_z \ dt
\end{equation}
\noindent where $e_z = \omega_{z,ref} - \hat\omega_z $ is the error between the desired and actual yaw rates, and $k_p$ and $k_i$ are the proportional and integral gains, respectively. A first-order low-pass filter ($\tau = 0.4$ s) is applied to the measured yaw rate $\hat{\boldsymbol \omega}_z$ for high-frequency noise attenuation.

For testing and integration, an angular control capability is also implemented by cascading an outer PI loop on top of the speed controller. In this mode, the angle error is converted into a commanded yaw rate, which replaces the nominal $\omega_{z,ref}$ input of the inner loop.


\section{Numerical Results}
\vspace{6pt}
Numerical simulations are performed using the process and observation models described in the previous section. The dynamic simulation assumes a rigid-body Taurus gondola with mass of $m=826$ kg, inertia matrix at the pivot resolved in $\mathcal{F}_b$ of $\mathbf{I} = [\,3.8{\times}10^{3},\ 1.4,\ -1.6;\ 1.4,\ 3.8{\times}10^{3},\ -5.1;\ -1.6,\ -5.1,\ 3.4{\times}10^{2}\,] \,\text{kg}\cdot\text{m}^2$, and a pivot-to-COM distance of $L = 1.94$ m, the former measured from the experimental setup and the latter two taken from the Computer Aided Design (CAD) model. The system damping matrix is set to $\mathbf{D}=\operatorname{diag}(\ 200, \ 200,\ 0\ )\ \mathrm{Nm\cdot s/rad}$, and the Coulomb torque parameters to $c_z=0.75$ and $\omega_\epsilon =10^{-2}$. The desired gondola yaw angular velocity, from science requirements, is $\omega_{z, des}=\pm 30^\circ \mathrm{s^{-1}}$. A ramp function is defined with $\dot \omega_{z,des}=\pm 1^\circ s^{-2}$ to avoid discontinuities in the commanded yaw rate and large torque demands. Control gains are tuned to $k_p= 1\ \mathrm{Nm}\cdot \mathrm{s}/^\circ$  and $k_i=0.2\ \mathrm{Nm}\cdot \mathrm{s}^{-1}/^\circ$. The inertial reference vectors were chosen as $\mathbf y_i^1\in\{[0,0,1]^\top,\,[0,1,0]^\top\}$ for camera~1 and $\mathbf y_i^2\in\{\tfrac{1}{\sqrt{3}}[1,1,1]^\top,\,\tfrac{1}{\sqrt{2}}[1,-1,0]^\top\}$ for camera~2. Simulation measurement and process noise parameters are presented in Table \ref{tab1}. All simulations are performed with a time step of $\Delta t = 0.001$ s using a second-order Runge–Kutta (midpoint) integration scheme.

\input{tables/tab1}

Results for the simulated yaw-rate control are shown in Fig.~\ref{fig:4}, which for now assumes a perfect tuning of the attitude estimator. Starting from rest, the simple PI torque controller tracks the commanded yaw rate with overshoot and settling time within specifications. Steady-state error is negligible once the integral term converges. At spin-up, the pivot torque gives a short positive pulse, then settles to a small near-zero bias during the constant-rate segment to balance viscous/Coulomb losses. A symmetric negative torque command also occurs at spin-down. The close-up of the steady-state tracking region shows small residual fluctuations around the commanded yaw rate of $30^\circ$/s. A batch of 20 Monte Carlo simulations was performed for this controlled run, and results for selected performance metrics are summarized in Table \ref{tab2}. The reported settling time is defined relative to the commanded yaw-rate ramp rather than a step input. We find that higher noise mainly increases the steady-state yaw-rate standard deviation, whereas overshoot and settling time are typically set by the closed-loop dynamics and controller gains. Overall, although the control performance is not aggressive, it remains well within Taurus’ science requirements, which place greater emphasis on accurate attitude estimation than on precise control.

\input{figures/fig4}
\input{tables/tab2}

\subsection{MEKF Performance}
To numerically evaluate MEKF performance, we simulate a free-response case in which the gondola is impulsively disturbed and then allowed to decay with no motor control, similar to Fig \ref{fig:2}. Attitude data is generated using
Eq. (\ref{eq:Euler}) initialized at $\boldsymbol\omega_0=[\ -0.5\ \ 0.5\ \ -10.0\ ]^{\mathsf{T}} \  ^\circ/\mathrm{s}$, $\mathbf{C}_{bi}(0) = \mathbf{1}$, and $\mathbf{b}_0=[\ 0.05\ \ 0.03\ \ -0.06\ ]^{\mathsf{T}} \ ^\circ/\mathrm{s}$. Table~\ref{tab3} lists the MEKF initialization and tuning quantities, including the initial covariance $\hat{\mathbf{P}}_0$, the noise covariances $\mathbf{Q}$ and $\mathbf{R}$, and the initial estimates $\hat{\mathbf{C}}_{bi,0}$ and $\hat{\mathbf{b}}_0$. We initialize the attitude with a few degrees of error to demonstrate the MEKF’s performance and slightly overestimate the covariances $\mathbf{Q}$ and $\mathbf{R}$.

\input{tables/tab3}

The attitude estimation error is defined from the relative rotation between the estimated and true attitude matrices as
\begin{equation}
    \mathbf{C}_{\text{err}} = \hat{\mathbf{C}}_{bi}\,\mathbf{C}_{bi}^{\mathsf{T}},
    \qquad
    \delta\boldsymbol{\theta} =  \bigl(\log\bigl(\mathbf{C}_{\text{err}}\bigr)\bigr)^\otimes \in \mathbb{R}^3.
\end{equation}

\noindent where $\mathbf{C}_{\text{err}} = \mathbf{1}$ when the estimate matches the true DCM exactly.  The corresponding scalar attitude error angle is
\begin{equation}
    \|\delta\boldsymbol{\theta}\|
    = \cos^{-1}\!\left(\frac{\operatorname{tr}(\mathbf{C}_{\text{err}})-1}{2}\right).
\end{equation}

\begin{figure}[t]
  \centering
  \includegraphics[width=0.7\linewidth]{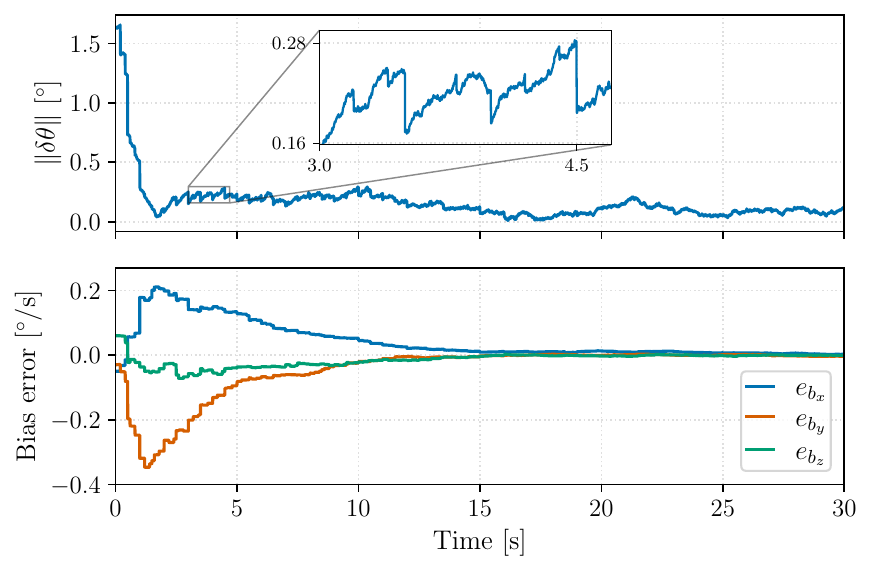}
  \caption{Attitude and bias estimation errors for the MEKF simulation with low-noise parameters.}
  \label{fig6}
\end{figure}

Results for the attitude and bias errors for a single realization are shown in Fig.~\ref{fig6}. As before, 20 Monte Carlo runs were carried out to obtain the steady-state statistics presented in Table \ref{tab4}. The attitude error converges rapidly below $0.3^\circ$ followed by a small steady-state residual oscillation centered at $0.0745 \pm 0.0058 ^\circ$. The zoomed inset shows how the attitude estimate is periodically reduced by vector measurements between short stretches of gyro dead-reckoning drift. The bias error in the lower plot is found to converge to a near-zero steady-state value. The small residual bias error (Table \ref{tab4}) is consistent with the presence of process disturbances (e.g. $\boldsymbol\tau_{ noise},\ \delta\boldsymbol\xi_k$), which result in a small steady-state offset in the estimated bias. The plot illustrates a trade-off in the MEKF between fast attitude convergence and accurate bias estimation. Initially, with large attitude error and limited vector-measurement authority, the filter attributes part of the mismatch between predicted and observed dynamics to the gyro biases, causing the bias states to jump as they temporarily absorb the error. This allows for rapid reduction of $\|\delta\boldsymbol{\theta}\|$ within the first few seconds, but at the expense of transient bias accuracy. As vector updates accumulate and the attitude estimate becomes well constrained, the filter can better distinguish true angular motion from bias drift, leading the bias errors to decay smoothly toward zero over a longer timescale. Overall, the MEKF behaves as expected and yields stable attitude and bias estimates under realistic operating conditions.

\input{tables/tab4}

\section{Experimental Results}
\vspace{6pt}
Preliminary experiments were carried out using the Taurus gondola with a mass proxy in place of the cryostat. The payload was suspended from the ceiling using straps that connect to the universal joint. The PI control strategy was implemented to evaluate basic yaw-rate tracking and system response. Since no cameras were installed during these tests, the MEKF was not implemented, and only gyro-based yaw rate control was evaluated. A Raspberry Pi 4 interfaces with a KVH DSP-1760 fiber-optic gyroscope sampling at 100 Hz, and an AMC DPRALTR motor driver controlling a Parker K178200-8Y1-CE frameless servo motor for pivot actuation. The mass of the system was measured at $826$ kg. PI control gains were tuned experimentally to $k_p=0.5\ \mathrm{Nm}\cdot \mathrm{s}/^\circ$ and $k_i = 0.1\ \mathrm{Nm}\cdot \mathrm{s}^{-1}/^\circ$. The experimental setup is shown in Fig. \ref{fig:3}, along with representative free-response dynamics from an initial impulse to the system. A simple qualitative comparison of Fig.~\ref{fig:3}b and Fig.~\ref{fig:2}b supports that the dynamic model (\ref{eq:Euler}) generally captures the behavior observed in the experimental results.

\input{figures/fig3}

\subsection{Yaw Rate Control}

Step-response experiments were conducted to assess the ability of the pivot actuation system to regulate yaw rate using the PI controller described in Section~\ref{sec:control}. Five ramp step commands were executed, alternating between $\pm 30^\circ/\mathrm{s}$ references. The corresponding performance metrics are summarized in Table \ref{tab5}.

Across all runs, the system achieved the commanded yaw-rate profiles with consistent behavior. The spin-up segments showed a small overshoot before settling to the target rate, and the return-to-zero transitions followed the same pattern. The zoom in Fig.~\ref{fig5} shows a $\sim0.8$ Hz yaw oscillation caused by the strap suspension acting like a spring in the test setup, which is specific to this configuration and may be different on the actual flight system. Overall, the experimental results were found to be in agreement with the numerical predictions, with comparable overshoot, somewhat longer settling times, and reduced steady-state yaw-rate noise, as shown in Table \ref{tab5}. The longer settling time is reasonable given hardware and torsional flight train effects that are not fully represented in the simplified model. The lower steady state noise may be due either to unmodeled mechanical damping or to more conservative noise assumptions in simulation. Most importantly, the results confirm that pivot-only yaw-rate control performs as expected on the real system and that the model framework captures the essential dynamics. These insights will guide future development of model-based control strategies, including the integration of star camera measurements in the experimental setup and evaluation of feedforward compensation.

\input{figures/fig5}

\input{tables/tab5}

\section{Conclusion}
\vspace{6pt}

In this paper, we introduced the yaw-rate control problem for balloon-borne payloads with pivot-only actuation through the lens of the Taurus experiment. We implemented a simple PI controller and a Multiplicative Extended Kalman Filter for attitude estimation and control. The approach was validated using numerical simulations with representative disturbance and noise levels, as well as preliminary experimental tests on the Taurus gondola. The experimental results demonstrated reliable high-rate tracking, robust estimator performance, and close agreement with the simulated behavior, indicating that the simplified model captures the essential dynamics relevant to controller design. Future work includes developing a more advanced controller to improve transient performance, experimentally implementing the MEKF with onboard inertial and optical sensor data, and integrating the $\tau$HK housekeeping system \cite{Tartakovsky2025} to meet the sensor and thermal requirements of the flight payload.

\section*{Appendix}
\subsection{Data Post-Processing}
\subsubsection{Sensor Alignment}

The yaw-axis gyroscope was aligned using a set of controlled ``ballerina twirl'' experiments, in which the gondola was commanded to spin about its yaw axis at constant speed. 
Assuming negligible pitch and roll motion, the aligned angular velocity is
\begin{equation}
    \boldsymbol{\omega}_{\mathrm{al}} = [\,0\ 0\ \|\boldsymbol{\omega}_{\mathrm{meas}}\|_2\,]^{\mathsf{T}} .
\end{equation}
A small, constant misalignment between the sensor and body frames is modeled as
\begin{equation}
    \boldsymbol{\omega}_{\mathrm{al}} = \mathbf{C}_{\mathrm{al}}\,\boldsymbol{\omega}_{\mathrm{meas}}, 
    \qquad 
    \mathbf{C}_{\mathrm{al}} \approx (\mathbf{1} - \delta\boldsymbol{\theta}^\times),
\end{equation}
with $\delta\boldsymbol{\theta}$ as the misalignment vector. Substituting gives
\begin{equation}
    -\boldsymbol{\omega}_{\mathrm{meas}}^{\times}\,\delta\boldsymbol{\theta} 
    = \boldsymbol{\omega}_{\mathrm{al}} - \boldsymbol{\omega}_{\mathrm{meas}} .
\end{equation}
Stacking samples from each twirl run, we solve the least-squares system
\begin{equation}
\mathbf{A}\,\delta\boldsymbol{\theta} = \mathbf{b},
    \qquad 
    \hat{\delta\boldsymbol{\theta}} = (\mathbf{A}^{\mathsf{T}}\mathbf{A})^{-1}\mathbf{A}^{\mathsf{T}}\mathbf{b},
\end{equation}
to estimate the sensor misalignment. For example, the resulting alignment vector for the experimental setup is 
$\hat{\delta\boldsymbol{\theta}} = [-0.447 \pm 0.003,\,-1.095 \pm 0.004,\,0.0000 \pm 0.0000 ]^\circ$.

\subsubsection{Noise \& Bias Characterization}
A recorded run with no payload movement provides information on the gyroscope bias and noise levels. With no motion, the measured rates reflect sensor bias, white noise, and the small projection of Earth's rotation. The mean rate was taken as the bias estimate, and the residuals were used to compute the noise covariance, later used to tune the gyro noise and bias random-walk parameters in the estimator model.

\section*{Acknowledgments}
\vspace{6pt}
The authors acknowledge the work of StarSpec Technologies, responsible for design and assembly of the Taurus gondola. Taurus is supported in the USA by NASA award number 80NSSC21K1957, and is supported in Europe by The Icelandic Research Fund (Grant number: 2410656-051) and the European Union (ERC, CMBeam, 101040169). 

\bibliography{sample}

\end{document}

%% file: figures/fig2.tex
\begin{figure}[t!]
  \centering
  \begin{tikzpicture}

    \def\figXshift{-0cm}    
    \def\figYshift{-0cm}    

    \def\plotXshift{1.2cm}  
    \def\plotYshift{6.0cm}   

    \begin{scope}[shift={(\figXshift,\figYshift)}]
      \node[anchor=south west, inner sep=0] (img)
        {\input{figures/fig2.5}}; 

      \begin{scope}[shift={([xshift=\plotXshift,yshift=\plotYshift]img.south east)}]
        \begin{groupplot}[
          group style={group size=1 by 3, vertical sep=1.0em},
          width=0.53\textwidth,
          height=3.55cm,
          tick label style={font=\normalsize},
          label style={font=\normalsize},
          no markers,
          axis line style={line width=0.3pt, color=black},
          tick style={line width=0.4pt, color=black},
          tickwidth=2.5pt,
          ytick pos=left,
          xtick pos=bottom, 
          xtick align=outside,
          ytick align=outside,
          every y label/.style={rotate=90, yshift=-2.0em},
          every axis x label/.style={at={(0.5,-0.2)}, anchor=north},
          enlargelimits=false,
          clip=false
        ]

          \nextgroupplot[
            ylabel={Roll Rate [$^\circ$/s]},
            xticklabels={},
            ytick={-3,0,3},
            xmin=0, ymin=-3.5, ymax=3.5
          ]
            \addplot[color={rgb,255:red,0; green,158; blue,115}, line width=1.5pt, smooth]
              table [x={t [s]}, y={roll_rate [deg/s]}, col sep=comma]
              {data/free_decay_body_rates_model.csv};

          \nextgroupplot[
            ylabel={Pitch Rate [$^\circ$/s]},
            xticklabels={},
            ytick={-3,0,3},
            xmin=0, ymin=-4.5, ymax=4.5
          ]
            \addplot[color={rgb,255:red,230; green,159; blue,0}, line width=1.5pt, smooth]
              table [x={t [s]}, y={pitch_rate [deg/s]}, col sep=comma]
              {data/free_decay_body_rates_model.csv};

          \nextgroupplot[
            ylabel={Yaw Rate [$^\circ$/s]},
            xlabel={Time [s]},
            ytick={-10,0,10},
            xtick={0,25,50,75,100},
            xmin=0, ymin=-1, ymax=11
          ]
            \addplot[color={rgb,255:red,86; green,180; blue,233}, line width=1.5pt, smooth]
              table [x={t [s]}, y expr=-\thisrow{yaw_rate [deg/s]}, col sep=comma]
              {data/free_decay_body_rates_model.csv};

        \end{groupplot}
      \end{scope}
    \end{scope}

  \end{tikzpicture}

  \caption{Representative free-body diagram of the empty Taurus gondola with reference frames and external torque at the pivot (left), and simulated free-response dynamics from an initial $2^\circ$ tilt from the vertical axis with $\boldsymbol{\omega}_0 = [-0.5,\ \ 0.5,\ -10.0]^{\mathsf{T}} \  ^\circ/\mathrm{s}$ (right).}
  \label{fig:2}
\end{figure}

%% file: figures/fig2.5.tex
\begin{tikzpicture}
  \node[anchor=south west, inner sep=0] (img)
    {\includegraphics[width=0.45\linewidth,
      trim={1.5cm 0cm 0cm 0cm},clip]
      {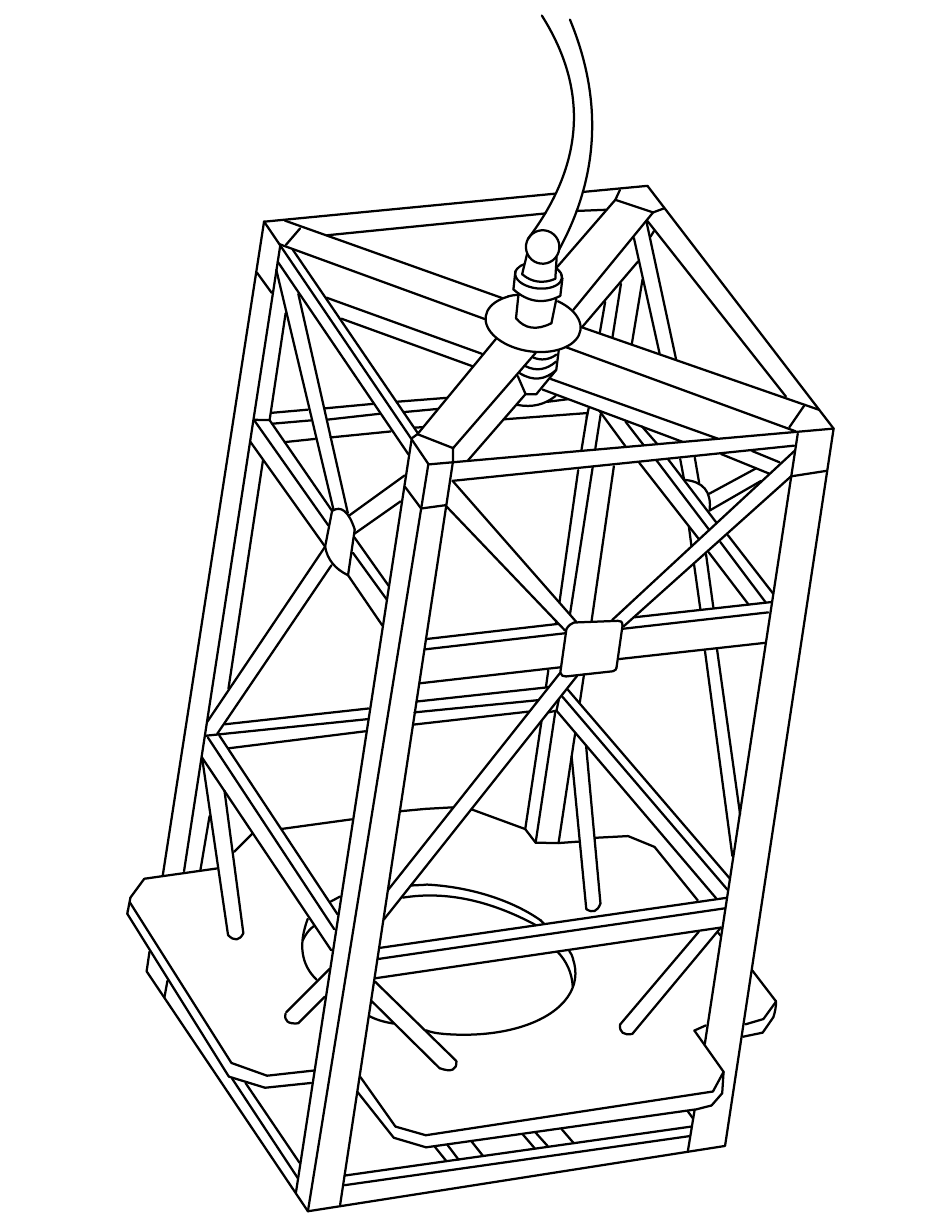}};

  \begin{scope}[x={(img.south east)}, y={(img.north west)}]

    \begin{scope}[shift={(0.505,0.50)}]
      \tikzset{axisarrow/.style={->, >=stealth, line width=3pt}}
      \draw[axisarrow] (0,0) -- ++({0.2*cos(184.5)},{0.2*sin(184.5)});    
      \draw[axisarrow] (0,0) -- ++({0.15*cos(80)},{0.15*sin(80)});        
      \draw[axisarrow] (0,0) -- ++({0.16*cos(-40)},{0.16*sin(-40)});      
      \node[font=\normalsize] at (-0.084,0.011) {$\mathcal{F}_b$};
    \end{scope}

    \begin{scope}[shift={(0.1,0.84)}]
      \def\L{0.08}
      \draw[->, >=stealth, line width=1.5pt] (0,0) -- (\L,0);
      \draw[->, >=stealth, line width=1.5pt] (0,0) -- (0,0.065);
      \draw[->, >=stealth, line width=1.5pt] (0,0) -- (-0.04,-0.04);
      \node[font=\normalsize] at (0.03,0.03) {$\mathcal{F}_i$};
    \end{scope}

    \draw[->, >=stealth, line width=1.0pt] (0.05,0.70) -- (0.05,0.62);
    \node[font=\normalsize] at (0.08,0.62) {$\mathbf{g}$};

  \def\ctrx{0.515}       
  \def\ctry{0.738}       
  \def\axisang{182}       
  \def\radius{0.08}      
  \def\tilt{0}          
  \def\arcstart{-50}     
  \def\arcend{230}        

\begin{scope}[rotate around={\axisang:(\ctrx,\ctry)}]
  \draw[line width=2pt, domain=\arcend:\arcstart, samples=100, smooth]
    plot ({\ctrx + \radius*cos(\x)}, {\ctry + 0.4*\radius*sin(\x+\tilt)});

  \path let
    \p1 = ({\ctrx + \radius*cos(\arcstart)}, {\ctry + 0.4*\radius*sin(\arcstart+\tilt)}),
    \p2 = ({\ctrx + \radius*cos(\arcstart-6)}, {\ctry + 0.4*\radius*sin(\arcstart-6+\tilt)})
  in
    (\p1) edge[
      -{Stealth[length=7pt,width=5pt]},
      line width=2pt,
      shorten >=-3pt,  
      line cap=round     
    ] (\p2);
\end{scope}

    \node[font=\normalsize] at (0.41,0.78) {$\boldsymbol \tau_{ext}$};    

  \end{scope}
\end{tikzpicture}

%% file: tables/tab1.tex
\begin{table}[h]
  \centering
  \renewcommand{\arraystretch}{1.2}
  \caption{Noise parameter values used in simulation.}
  \label{tab1}
  \begin{tabularx}{0.7\textwidth}{%
    >{\centering\arraybackslash}X|
    >{\centering\arraybackslash}X|
    >{\centering\arraybackslash}X}
    \hline
    Noise Parameter & Low-Noise Case & High-Noise Case \\
    \hline\hline
    $\boldsymbol\Sigma_\tau\ \mathrm{[Nm]^2}$& $\operatorname{diag}( 0.5, \ 0.5, \ 0.01 )\ \Delta t$& $\operatorname{diag}( 1.0, \ 1.0, \ 0.02 )\ \Delta t$ \\ 
    \hline
     $\boldsymbol\Sigma_\xi \ \mathrm{[deg]^2}$ &  $0.001^2 \cdot\mathbf{1}$ &$ 0.001^2 \cdot\mathbf{1}$\\ 
    \hline
     $\boldsymbol\Sigma_g\ \mathrm{[deg/s]^2}$& $0.02^2/\Delta t\cdot\mathbf{1}$ & $0.06^2/\Delta t\cdot\mathbf{1}$\\ 
    \hline
     $\boldsymbol\Sigma_b\ \mathrm{[deg/s]^2}$& $0.001^2\Delta t\cdot\mathbf{1}$ & $0.002^2\Delta t\cdot\mathbf{1}$ \\ 
    \hline
     $\boldsymbol\Sigma_{n,1}\ \mathrm{[deg]^2}$& $0.1^2\cdot \mathbf{1}$ & $0.5^2\cdot \mathbf{1}$\\ 
    \hline
    $\boldsymbol\Sigma_{n,2}\ \mathrm{[deg]^2}$& $0.2^2\cdot \mathbf{1}$ & $0.5^2 \cdot \mathbf{1}$ \\ 
    \hline
    $f_{\mathrm{sc},1}$ $\mathrm{[Hz]}$& $2.0$ & $0.5$ \\ 
    \hline
    $f_{\mathrm{sc},2}$ $\mathrm{[Hz]}$& $5.0$ & $0.5$ \\ 
    \hline

  \end{tabularx}
\end{table}

%% file: figures/fig4.tex
\begin{figure}[ht]
  \centering
  \begin{tikzpicture}
    \begin{groupplot}[
      group style={group size=1 by 2, vertical sep=1.0em},
      width=0.6\textwidth,
      height=4.2cm,
      grid=both,
      grid style={dotted, gray!50},
      tick label style={font=\normalsize},
      label style={font=\normalsize},
      axis line style={line width=0.3pt, color=black},
      tickwidth=2.5pt,
      ytick pos=left,
      xtick pos=bottom, 
      xtick align=outside,
      ytick align=outside,
      tick style={line width=0.4pt, color=black},
      clip=false
    ]

      \nextgroupplot[
        height=5.1cm,
        ylabel={Yaw rate [$^\circ$/s]},
        xticklabels={},
        xmin=0, xmax=360,
        legend style={
          at={(0.5,1.02)},
          anchor=south,
          draw=none,
          fill=none,
          font=\normalsize,
          legend columns=-1
        }
      ]
        \addplot[
          color={rgb,255:red,0; green,114; blue,178},
          line width=1.8pt
        ]
          table [x={time_s}, y={w_yaw_deg_s}, col sep=comma]
          {data/yaw_control_data_sim.csv};
        \addlegendentry{Measured}

        \addplot[
          dashed,
          color={rgb,255:red,230; green,159; blue,0},
          line width=1.8pt
        ]
          table [x={time_s}, y={w_ref_deg_s}, col sep=comma]
          {data/yaw_control_data_sim.csv};
        \addlegendentry{Reference}

        \coordinate (zoomSW) at (axis cs:190,28.5); 
        \coordinate (zoomNW) at (axis cs:190,31.5); 
        \coordinate (zoomNE) at (axis cs:220,31.5);
        \draw[draw=red!10!black] (zoomSW) rectangle (zoomNE);
        (axis cs:190,28.5) rectangle (axis cs:220,31.5);

      \nextgroupplot[
        height=3.7cm,
        ylabel={$\tau_{piv}$ [N·m]},
        ytick={-10, 0, 10},
        xlabel={Time [s]},
        xmin=0, xmax=360,
        ymin=-10, ymax=12
      ]
        \addplot[
          color={rgb,255:red,213; green,94; blue,0},
          line width=1.8pt
        ]
          table [x={time_s}, y={tau_z_Nm}, col sep=comma]
          {data/yaw_control_data_sim.csv};

    \end{groupplot}

    \begin{axis}[
        name=inset,
        width=4cm,
        height=1.4cm,
        at={(2.15cm,0.6cm)},    
        anchor=south west,
        scale only axis,
        xmin=190, xmax=220,  
        xtick={190,220},
        ytick={29.93,30,30.06},
        grid=both,
        tick align=outside,
        tick style={line width=0.4pt, color=black},
        xtick pos=bottom,
        ytick pos=left,
        xtick align=outside,
        ytick align=outside,
        tickwidth=2.5pt,
        grid style={dotted, gray!40},
        tick label style={font=\scriptsize},
        label style={font=\scriptsize},
        axis line style={line width=0.3pt},
    ]
      \addplot[
          color={rgb,255:red,0; green,114; blue,178},
          line width=0.8pt
      ]
        table [x={time_s}, y={yaw_rate_deg_s}, col sep=comma]
        {data/controlled_run_sim_inset.csv};

      \addplot[
        dashed,
        color={rgb,255:red,230; green,159; blue,0},
        line width=0.8pt
        ]
        coordinates {(190,30) (220,30)};
        
    \end{axis}

    \draw[draw=red!10!black] (zoomNE) -- (2.15cm+4cm,0.6cm+1.4cm);
    \draw[draw=red!10!black] (zoomNW) -- (2.15cm,0.6cm+1.4cm);
  \end{tikzpicture}

  \caption{One realization of simulated yaw-rate tracking (top) and corresponding control torque (bottom) for low-noise case.}
  \label{fig:4}
\end{figure}

%% file: tables/tab2.tex
\begin{table}[h]
  \centering
  \renewcommand{\arraystretch}{1.2}
  \caption{Monte Carlo performance statistics for yaw-rate control.}
  \label{tab2}
  \begin{tabularx}{0.7\textwidth}{l|Y|Y}
    \hline
    Metric & Low-Noise Case & High-Noise Case \\
    \hline\hline
    Overshoot [\%]                   & $11.14 \pm 0.12$      & $11.17 \pm 0.13$      \\ \hline
    Settling time (2\%) [s]          & $28.33 \pm 0.04$      & $28.34 \pm 0.06$      \\ \hline
    Steady-state yaw $1\sigma$ [deg/s] & $0.0306 \pm 0.0113$ & $0.0518 \pm 0.0175$   \\ \hline
  \end{tabularx}
\end{table}

%% file: tables/tab3.tex
\begin{table}[h]
  \centering
  \renewcommand{\arraystretch}{1.2}
  \caption{MEKF tuning parameter values used in simulation.}
  \label{tab3}
  \begin{tabularx}{0.6\textwidth}{>{\centering\arraybackslash}m{0.25\textwidth}|>{\centering\arraybackslash}X}
    \hline
    Parameter & Value \\
    \hline\hline
    $\hat{\textbf{P}}_0$ & $\operatorname{diag}\!\big((3^\circ)^2\mathbf{1},\ (0.07^\circ/\mathrm{s})^2\mathbf{1}\big)$ \\ 
    \hline
    $\textbf{Q}$ & $1.05\cdot\operatorname{diag}(\boldsymbol{\Sigma}_{g},\boldsymbol{\Sigma}_{b})$ \\ 
    \hline
    $\textbf{R}$ & $1.05\cdot\operatorname{diag}(\boldsymbol{\Sigma}_{n,1},\boldsymbol{\Sigma}_{n,2})$ \\ 
    \hline
    $\hat{\textbf{C}}_{bi,0}$ & $\exp(10\cdot[ 1\ \ 1\ \ 1 ]^\times)\cdot \mathbf{1}$ \\ 
    \hline
    $\hat{\textbf{b}}_0\ [^\circ/\mathrm{s}]$ & $\mathbf{0}$ \\ 
    \hline
  \end{tabularx}
\end{table}

%% file: tables/tab4.tex
\begin{table}[h]
  \centering
  \renewcommand{\arraystretch}{1.2}
  \caption{Monte Carlo steady-state performance statistics for the MEKF.}
  \label{tab4}
  \begin{tabularx}{0.95\textwidth}{l|Y}
    \hline
    Steady-state Metric ($t>15\mathrm{s}$) & Value \\ \hline\hline
    $\|\delta\boldsymbol\theta\|$ mean [$^\circ$]     
      & $0.0745 \pm 0.0058$ \\ \hline

    $\|\delta\boldsymbol\theta\|$ standard deviation ($1\sigma$) [$^\circ$]            
      & $0.0309 \pm 0.0025$ \\ \hline

    $(\hat{\mathbf{b}}-\mathbf{b})$ mean [$^\circ$/s]         
      & $[\,0.00270,\; -0.00097,\; -0.00171\,] 
         \pm [\,0.00362,\; 0.00421,\; 0.00396\,]$ \\ \hline

    $(\hat{\mathbf{b}}-\mathbf{b})$ standard deviation ($1\sigma$) [$^\circ$/s]                
      & $[\,0.00650,\; 0.00587,\; 0.00531\,] 
         \pm [\,0.00171,\; 0.00207,\; 0.00173\,]$ \\ \hline
  \end{tabularx}
\end{table}


%% file: figures/fig3.tex
\begin{figure}[t!]
  \centering
  \begin{tikzpicture}

    \def\figXshift{0cm}
    \def\figYshift{0cm}

    \def\plotXshift{2.0cm}
    \def\plotYshift{6.0cm}

    \begin{scope}[shift={(\figXshift,\figYshift)}]
      \node[anchor=south west, inner sep=0] (img)
  {\includegraphics[width=0.35\textwidth]{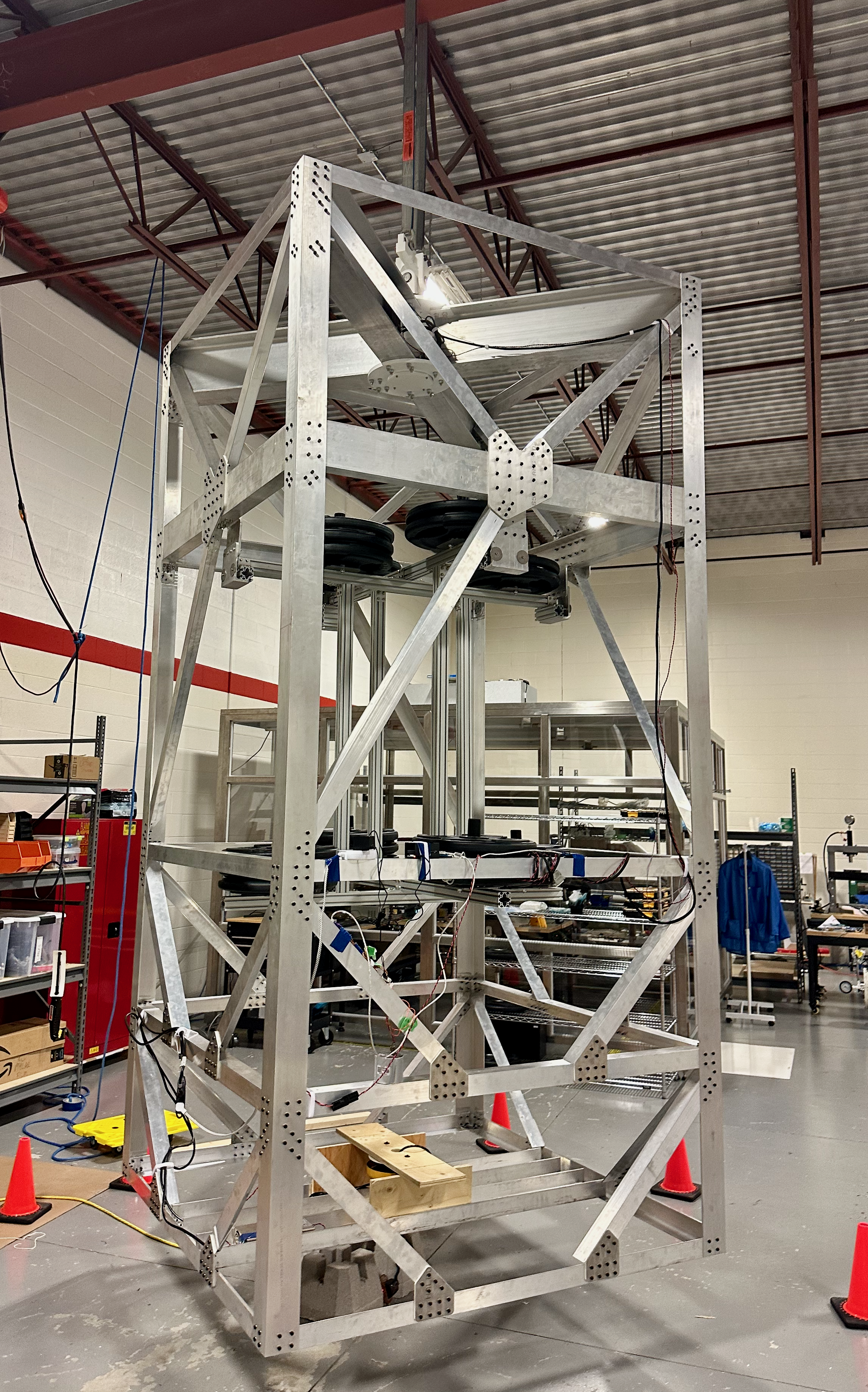}};
    \end{scope}

    \begin{scope}[shift={([xshift=\plotXshift,yshift=\plotYshift]img.south east)}]
      \begin{groupplot}[
        group style={group size=1 by 3, vertical sep=1.0em},
        width=0.55\textwidth,
        height=3.55cm,
        tick label style={font=\normalsize},
        label style={font=\normalsize},
        no markers,
        axis line style={line width=0.3pt, color=black},
        tick style={line width=0.4pt, color=black},
        tickwidth=2.5pt,
        ytick pos=left,
        xtick pos=bottom, 
        xtick align=outside,
        ytick align=outside,
        every y label/.style={rotate=90, yshift=-2.0em},
        every axis x label/.style={at={(0.5,-0.2)}, anchor=north},
        enlargelimits=false,
        clip=false
      ]

        \nextgroupplot[
          ylabel={Roll Rate [$^\circ$/s]},
          xticklabels={},
          ytick={-2,0,2},
          xmin=-0.1, ymin=-2.4, ymax=2.6
        ]
          \addplot[color={rgb,255:red,0; green,158; blue,115}, line width=1.5pt, smooth]
            table [x={time_s}, y={roll_rate_filt_deg_s}, col sep=comma, restrict x to domain=0:120]
            {data/lpfiltered_run3.csv};

        \nextgroupplot[
          ylabel={Pitch Rate [$^\circ$/s]},
          xticklabels={},
          ytick={-3,0,3},
          xmin=0, ymin=-3.5, ymax=3.5
        ]
          \addplot[color={rgb,255:red,230; green,159; blue,0}, line width=1.5pt, smooth]
            table [x={time_s}, y={pitch_rate_filt_deg_s}, col sep=comma, restrict x to domain=0:120]
            {data/lpfiltered_run3.csv};

        \nextgroupplot[
          ylabel={Yaw Rate [$^\circ$/s]},
          xlabel={Time [s]},
          ytick={-10,0,10},
          xtick={0,30, 60, 90,120},
          xmin=0, ymin=-1, ymax=10
        ]
          \addplot[color={rgb,255:red,86; green,180; blue,233}, line width=1.5pt, smooth]
            table [x={time_s}, y expr=-\thisrow{yaw_rate_filt_deg_s}, col sep=comma, restrict x to domain=0:120]
            {data/lpfiltered_run3.csv};

      \end{groupplot}
    \end{scope}

  \end{tikzpicture}

  \caption{Experimental setup of the suspended Taurus gondola in the StarSpec Technologies facilities (left), and  representative low-pass–filtered gyroscope angular-rate measurements illustrating free-response motion following an initial impulse (right).}
  \label{fig:3}
\end{figure}

%% file: figures/fig5.tex
\begin{figure}[ht]
  \centering
  \begin{tikzpicture}
    \begin{groupplot}[
      group style={group size=1 by 2, vertical sep=1.0em},
      width=0.6\textwidth,
      height=4.2cm,
      grid=both,
      grid style={dotted, gray!50},
      tick label style={font=\normalsize},
      label style={font=\normalsize},
      axis line style={line width=0.3pt, color=black},
      tickwidth=2.5pt,
      ytick pos=left,
      xtick pos=bottom, 
      xtick align=outside,
      ytick align=outside,
      tick style={line width=0.4pt, color=black},
      clip=false
    ]

      \nextgroupplot[
        height=5.1cm,
        ylabel={Yaw rate [$^\circ$/s]},
        xticklabels={},
        xmin=0, xmax=370,
        legend style={
          at={(0.5,1.02)},
          anchor=south,
          draw=none,
          fill=none,
          font=\normalsize,
          legend columns=-1
        }
      ]
        \addplot[
          color={rgb,255:red,0; green,114; blue,178},
          line width=1.8pt
        ]
          table [x={time_s}, y={yaw_deg_s}, col sep=comma]
          {data/controlled_run_downsampled.csv};
        \addlegendentry{Measured}

        \addplot[
          dashed,
          color={rgb,255:red,230; green,159; blue,0},
          line width=1.8pt
        ]
          table [x={time_s}, y={yaw_cmd_deg_s}, col sep=comma]
          {data/controlled_run_downsampled.csv};
        \addlegendentry{Reference}

        \coordinate (zoomSW) at (axis cs:140,28.5); 
        \coordinate (zoomNW) at (axis cs:140,31.5); 
        \coordinate (zoomNE) at (axis cs:170,31.5);
        \draw[draw=red!10!black] (zoomSW) rectangle (zoomNE);
        (axis cs:140,28.5) rectangle (axis cs:170,31.5);

      \nextgroupplot[
        height=3.7cm,
        ylabel={$\tau_{piv}$ [N·m]},
        ytick={-4, 0, 4},
        xlabel={Time [s]},
        xmin=0, xmax=370
      ]
        \addplot[
          color={rgb,255:red,213; green,94; blue,0},
          line width=1.8pt
        ]
          table [x={time_s}, y={pivot_torque_Nm}, col sep=comma]
          {data/controlled_run_downsampled.csv};

    \end{groupplot}

    \begin{axis}[
        name=inset,
        width=4cm,
        height=1.4cm,
        at={(2.15cm,0.6cm)},    
        anchor=south west,
        scale only axis,
        xmin=140, xmax=170,  
        xtick={140,170},
        ytick={29.96,30,30.04},
        grid=both,
        tick align=outside,
        tick style={line width=0.4pt, color=black},
        xtick pos=bottom,
        ytick pos=left,
        xtick align=outside,
        ytick align=outside,
        tickwidth=2.5pt,
        grid style={dotted, gray!40},
        tick label style={font=\scriptsize},
        label style={font=\scriptsize},
        axis line style={line width=0.3pt},
    ]
      \addplot[
          color={rgb,255:red,0; green,114; blue,178},
          line width=0.8pt
      ]
        table [x={time_s}, y={yaw_deg_s}, col sep=comma]
        {data/controlled_run_inset.csv};

      \addplot[
        dashed,
        color={rgb,255:red,230; green,159; blue,0},
        line width=0.8pt
        ]
        coordinates {(140,30) (180,30)};
        
    \end{axis}

    \draw[draw=red!10!black] (zoomNE) -- (2.15cm+4cm,0.6cm+1.4cm);
    \draw[draw=red!10!black] (zoomNW) -- (2.15cm,0.6cm+1.4cm);
  \end{tikzpicture}

  \caption{One realization of experimental yaw-rate tracking (top) and corresponding control torque (bottom).}
  \label{fig5}
\end{figure}
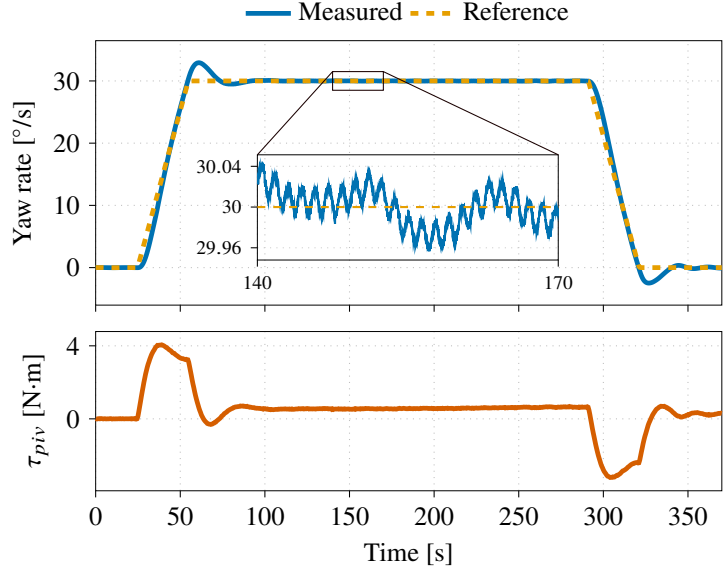

%% file: tables/tab5.tex
\begin{table}[h]
  \centering
  \renewcommand{\arraystretch}{1.2}
  \caption{Experimental performance statistics for yaw-rate control from 5 controlled runs.}
  \begin{tabularx}{0.55\textwidth}{l|Y}
    \hline
    Metric & Value \\
    \hline\hline
    Overshoot [\%]                     & $9.60 \pm 0.69$      \\ \hline
    Settling time (2\%) [s]            & $41.715 \pm 0.414$      \\ \hline
    Steady-state yaw $1\sigma$ [deg/s] & $0.0187 \pm 0.0009$   \\ \hline
  \end{tabularx}
  \label{tab5}
\end{table}

%% file: sample.bib
@article{Adler2024,
  title        = {Modeling optical systematics for the Taurus CMB experiment},
  author       = {Adler, Alexandre E. and Austermann, Jason E. and Benton, Steven J. and Duff, Shannon M. and Filippini, Jeffrey P. and Fraisse, Aurelien A. and Gascard, Thomas and Gibbs, Sho M. and Gourapura, Suren and Hubmayr, Johannes and Gudmundsson, Jon E. and Jones, William C. and May, Jared L. and Nagy, Johanna M. and Okun, Kate and Padilla, Ivan and Rooney, Christopher and Tartakovsky, Simon and Vissers, Michael R.},
  journal      = {Journal of Cosmology and Astroparticle Physics},
  volume       = {2024},
  number       = {09},
  pages        = {061},
  year         = {2024},
  doi          = {10.1088/1475-7516/2024/09/061},
  eprint       = {2406.11992}
}

@book{Barfoot2024,
  author    = {Timothy D. Barfoot},
  title     = {State Estimation for Robotics},
  edition   = {2nd},
  publisher = {Cambridge University Press},
  year      = {2024},
  isbn      = {9781009299909}
}

@article{Gill2024,
  author = {Gill, Ajay S. and Benton, Steven J. and Damaren, Christopher J. and Everett, Spencer W. and Fraisse, Aurelien A. and Hartley, John W. and Harvey, David and Holder, Bradley and Huff, Eric M. and Jauzac, Mathilde and Jones, William C. and Lagattuta, David and Leung, Jason S.-Y. and Li, Lun and Luu, Thuy Vy T. and Massey, Richard and McCleary, Jacqueline E. and Nagy, Johanna M. and Netterfield, C. Barth and Paracha, Emaad and Redmond, Susan F. and Rhodes, Jason D. and Robertson, Andrew and Romualdez, L. Javier and Schmoll, Jürgen and Shaaban, Mohamed M. and Sirks, Ellen L. and Vassilakis, Georgios N. and Vitorelli, André Z.},
  title = {SuperBIT Superpressure Flight Instrument Overview and Performance: Near-diffraction-limited Astronomical Imaging from the Stratosphere},
  journal = {The Astronomical Journal},
  volume = {168},
  number = {2},
  pages = {85},
  year = {2024},
  publisher = {American Astronomical Society},
  doi = {10.3847/1538-3881/ad5840}
}

@article{Crassidis2007,
  author    = {John L. Crassidis and F. Landis Markley and Yang Cheng},
  title     = {Survey of Nonlinear Attitude Estimation Methods},
  journal   = {Journal of Guidance, Control, and Dynamics},
  year      = {2007},
  volume    = {30},
  number    = {1},
  pages     = {12--28},
  month     = {January},
  doi       = {10.2514/1.22452},
  publisher = {American Institute of Aeronautics and Astronautics}
}

@article{Bernasconi2025,
  title        = {The Gondola for the SUNRISE III Balloon-Borne Solar Observatory},
  author       = {Bernasconi, Pietro and Carpenter, Michael and Eaton, Harry and Schulze, Erich and Carkhuff, Bliss and Palo, Geoffrey and Young, Daniel and Raouafi, Nour and Vourlidas, Angelos and Coker, Robert and Solanki, Sami K. and Korpi-Lagg, Andreas and Gandorfer, Achim and Feller, Alex and Riethmüller, Tino L. and Smitha, H. N. and Grauf, Bianca and del Toro Iniesta, Jose Carlos and Orozco Suárez, David and Katsukawa, Yukio and Kubo, Masahito and Berkefeld, Thomas and Bell, Alexander and {\'A}lvarez-Herrero, Alberto and Mart{\'i}nez Pillet, Valent{\'i}n},
  journal      = {Solar Physics},
  volume       = {300},
  number       = {112},
  year         = {2025},
  month        = aug,
  publisher    = {Springer},
  doi          = {10.1007/s11207-025-02490-1}
}

@article{Cui2025,
  author       = {Cui, Yulang and Zhou, Jianghua and Li, Yijian and Huang, Wanning and Liu, Yongqi},
  title        = {Multi-Stage Coordinated Azimuth Control for High-Precision Balloon-Borne Astronomical Platforms},
  journal      = {Aerospace},
  volume       = {12},
  number       = {9},
  pages        = {821},
  year         = {2025},
  month        = sep,
  publisher    = {MDPI},
  doi          = {10.3390/aerospace12090821}
}

@article{Quadrelli2004,
  author       = {Quadrelli, Marco B. and Cameron, Jonathan M. and Kerzhanovich, Viktor},
  title        = {Multibody Dynamics of Parachute and Balloon Flight Systems for Planetary Exploration},
  journal      = {Journal of Guidance, Control, and Dynamics},
  volume       = {27},
  number       = {4},
  pages        = {647--659},
  year         = {2004},
  month        = jul,
  doi          = {10.2514/1.11337}
}

@article{Du2024,
  author    = {Jingyuan Du and Xingguo Wei and Jian Li and Gangyi Wang and Xiaowei Wan},
  title     = {Star Spot Extraction for Multi-FOV Star Sensors Under Extremely High Dynamic Conditions},
  journal   = {IEEE Sensors Journal},
  volume    = {24},
  number    = {21},
  pages     = {35167--35180},
  year      = {2024},
  month     = {Nov},
  doi       = {10.1109/JSEN.2024.3459001}
}

@article{Pascale2008,
  author       = {Pascale, Enzo and Ade, P. A. R. and Bock, J. J. and Chapin, E. L. and Chung, J. and Devlin, M. J. and Dicker, S. and Griffin, M. and Gundersen, J. O. and Halpern, M. and Hargrave, P. C. and Hughes, D. H. and Klein, J. and MacTavish, C. J. and Marsden, G. and Martin, P. G. and Martin, T. G. and Mauskopf, P. and Netterfield, C. B. and Olmi, L. and Patanchon, G. and Rex, M. and Scott, D. and Semisch, C. and Thomas, N. and Truch, M. D. P. and Tucker, C. and Tucker, G. S. and Viero, M. P. and Wiebe, D. V.},
  title        = {{The Balloon-borne Large Aperture Submillimeter Telescope: BLAST}},
  journal      = {The Astrophysical Journal},
  year         = {2008},
  volume       = {681},
  number       = {1},
  pages        = {400},
  doi          = {10.1086/588720},
}

@article{Tartakovsky2025,
  author       = {Tartakovsky, S. and Benton, S. J. and Fraisse, A. A. and Jones, W. C. and May, J. L. and Nagy, J. M. and Rodriguez, R. R. and Voyer, P.},
  title        = {$\tau$HK: A modular housekeeping system for cryostats and balloon payloads},
  journal      = {Review of Scientific Instruments},
  volume       = {96},
  number       = {9},
  pages        = {094902},
  year         = {2025},
  month        = {Sep},
  doi          = {10.1063/5.0282145}
}

@phdthesis{Romualdez2018,
    author = {Romualdez, Luis Javier},
    school = {University of Toronto},
    title = {{Design, Implementation, and Operational Methodologies for Sub-Arcsecond Attitude Determination, Control, and Stabilization of the Super-pressure Balloon-borne Imaging Telescope (SuperBIT)}},
    year = {2018}
    }

@inproceedings{May2024,
  author    = {J. L. May and A. E. Adler and J. E. Austermann and S. J. Benton and R. Bihary and M. Durkin and others},
  title     = {Instrument overview of Taurus: a balloon-borne CMB and dust polarization experiment},
  booktitle = {SPIE Ground-based and Airborne Telescopes X},
  volume    = {13094},
  pages     = {1296--1310},
  year      = {2024}
}

@inproceedings{Tartakovsky2024,
  author    = {S. Tartakovsky and A. E. Adler and J. E. Austermann and S. J. Benton and R. Bihary and M. Durkin and others},
  title     = {Thermal architecture for a cryogenic super-pressure balloon payload: design and development of the Taurus flight cryostat},
  booktitle = {SPIE Ground-based and Airborne Telescopes X},
  volume    = {13094},
  pages     = {1693--1701},
  year      = {2024}
}

@inproceedings{Nagler2022,
  author       = {Nagler, Peter C. and Bernard, Lee and Bocchieri, Andrea and Butler, Nat and Changeat, Quentin and D'Alessandro, Azzurra and Edwards, Billy and Gamaunt, John and Gong, Qian and Hartley, John and Helson, Kyle and Jensen, Logen and Kelly, Daniel P. and Klangboonkrong, Kanchita and Kleyheeg, Annalies and Lewis, Nikole K. and Li, Steven and Line, Michael and Maher, Stephen F. and McClelland, Ryan and Miko, Laddawan R. and Mugnai, Lorenzo V. and Netterfield, C. Barth and Parmentier, Vivien and Pascale, Enzo and Patience, Jennifer and Rehm, Tim and Romualdez, Javier and Sarkar, Subhajit and Scowen, Paul A. and Tucker, Gregory S. and Waczynski, Augustyn and Waldmann, Ingo},
  title        = {The EXoplanet Climate Infrared TElescope (EXCITE)},
  booktitle    = {Ground-based and Airborne Instrumentation for Astronomy IX},
  editor       = {Evans, Christopher J. and Bryant, Julia J. and Motohara, Kentaro},
  series       = {Proceedings of SPIE},
  volume       = {12184},
  pages        = {121840V},
  year         = {2022},
  organization = {International Society for Optics and Photonics},
  publisher    = {SPIE},
  doi          = {10.1117/12.2629373}
}

@article{Kassarian2024,
  author    = {E. Kassarian and F. Sanfedino and D. Alazard and J. Montel and C.-A. Chevrier},
  title     = {Modeling of stratospheric balloons and robust line-of-sight pointing control},
  journal   = {CEAS Space Journal},
  volume    = {16},
  pages     = {457--474},
  year      = {2024}
}

@book{deRuiter2012,
  author    = {A. H. de Ruiter and C. J. Damaren and J. R. Forbes},
  title     = {Spacecraft Dynamics and Control: An Introduction},
  publisher = {John Wiley \& Sons},
  year      = {2012}
}

@inproceedings{Shariff2014,
  author    = {J. A. Shariff and P. A. R. Ade and M. Amiri and S. J. Benton and J. J. Bock and J. R. Bond and S. A. Bryan and H. C. Chiang and C. R. Contaldi and B. P. Crill and O. P. Doré and M. Farhang and J. P. Filippini and L. M. Fissel and A. A. Fraisse and A. E. Gambrel and N. N. Gandilo and S. R. Golwala and J. E. Gudmundsson and M. Halpern and M. Hasselfield and G. C. Hilton and W. A. Holmes and V. V. Hristov and K. D. Irwin and W. C. Jones and Z. D. Kermish and C. L. Kuo and C. J. MacTavish and P. V. Mason and K. G. Megerian and L. Moncelsi and T. A. Morford and J. M. Nagy and C. B. Netterfield and R. O'Brient and A. S. Rahlin and C. D. Reintsema and J. E. Ruhl and M. C. Runyan and J. D. Soler and A. Trangsrud and C. E. Tucker and R. S. Tucker and A. D. Turner and C. E. Weber and D. V. Wiebe and E. Y. Young},
  title     = {Pointing control for the SPIDER balloon-borne telescope},
  booktitle = {Proceedings of SPIE - Ground-based and Airborne Telescopes V},
  volume    = {9145},
  pages     = {91450U},
  year      = {2014},
  doi       = {10.1117/12.2055166}
}
